\documentclass[preprint,showpacs,showkeys]{revtex4}
\usepackage[T2A]{fontenc}
\usepackage[cp1251]{inputenc}
\usepackage[english]{babel}
\usepackage{amsmath}
\usepackage{amsfonts}
\usepackage{graphicx}
\usepackage{epsfig}
\usepackage{subfigure}
\usepackage{amssymb}
\usepackage[indent,small]{caption2}

\begin{document}

\title{Anharmonic Bloch Oscillations in the Optical Waveguide Array}
\author{Gozman M. I.$^{1,2}$, Guseynov A. I.$^3$, Kagan Yu. M.$^1$, Pavlov A. I.$^1$, Polishchuk I. Ya.$^{1,2}$.}
\affiliation{$^1$ RRC Kurchatov Institute, Kurchatov Sq., 1, 123182 Moscow, Russia} %
\affiliation{$^2$ Moscow Institute of Physics and Technology, 141700, 9, Institutskii per., Dolgoprudny, Moscow Region, Russia} %
\affiliation{$^3$ Moscow Engineering Physics Institute, 115409, Kashirskoe highway, 31, Moscow, Russia} %

\begin{abstract}
The anharmonic Bloch oscillations of a light beam in the array of
optical waveguides are considered. The coupling modes model (CMM)
with the second order interaction is used to describe the effect
analytically. The formula obtained predicts explicitly the path of
the optical beam, in particular, the positions of the turning
points are found. A total agreement of this formula with the
numerical simulation is confirmed.
\end{abstract}

\maketitle


\section{Introduction}

\label{Sec:Intro}

Today the arrays of coupling optical waveguides attract the increasing
interest since they allow to control the behavior of the optical signal
effectively. Besides, such systems exhibit many interesting phenomena
similar to those that may occur in solids \cite{Review_Longhi2009}, such as
Bloch oscillations \cite{Pertsch1998_Theor, Pertsch1999, Morandotti1999,
Pertsch2002, Chiodo2006, Longhi2007, Gradons_Zheng2010, We_OptEng2014},
Zener tunneling \cite{Trompeter2006, Dreisow1_2009, Wang1_2010}, Anderson
localization \cite{AndLoc_Schwartz2008, AndLoc_Lahini2008}, dynamic
localization \cite{DinLoc_Garanovich2007, DinLoc_Dreisow2008} etc.

Usually, for the numerical simulations of the optical phenomena in the
arrays of interacting waveguides the coupling modes model (CMM) is used. This
model describes the behavior of light quite exactly if the coupling is
enough weak.

But the theoretical research can be considerably simplified if the numerical
simulation can be replaced with an analytical formula. In particular, for
the periodical arrays of the identical waveguides the equations of CMM can
be solved analytically \cite{Lederer2008}. The analytical
solutions were obtained for one-dimensional and two-dimensional arrays,
and the long-range interaction was taken into account.

Another problem which can be solved analytically is the array of waveguides
with the optical properties (such as radii or refractive indices) gradually
changing as one passes from one waveguide to another. This problem was
considered in \cite{Pertsch1998_Theor}. One of the main results of this work
is the analytical expression for the path of the optical beam. The obtained
formula predicts that the path of the optical beam takes the periodical
oscillating form. This analytical result was confirmed with the experiments
\cite{Pertsch1999, Morandotti1999, Pertsch2002, Chiodo2006}. This phenomenon
is known as optical Bloch oscillations. It is the optical
analog of the Bloch oscillations of an electron in an ordinary crystal
placed in an external electric field.

Generally, the optical Bloch oscillations, as well as the other optical
effects, are investigated for the plane arrays of the waveguides, since such
arrays are quite simple for fabrication. Besides, such systems are the
most convenient for the numerical simulation, since the coupling of the
adjacent waveguides only should be taken into account.

\begin{figure}[tbp]
\centering
\includegraphics[width=0.5\textwidth]{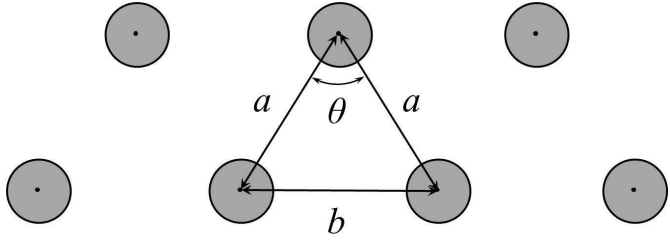}
\caption{The scheme of the zigzag array of waveguides.}
\label{Fig_Zigzag}
\end{figure}

However, today the increasing attention is drawn to the more complex
structures, namely zigzag arrays of waveguides (see Fig. \ref{Fig_Zigzag}).
Such arrays are interesting due to the significant role of the second order
coupling. In particular, in works \cite{Zigzag_Dreisow2008,
ZigzagSolitons_Efremidis2002, ZigzagSolitons_Szameit2009} the influence of
the second order coupling on the effects of diffraction and the formation of
solitons was investigated. Besides, recently the investigations of the
optical Bloch oscillations in zigzag arrays were published \cite{Wang2010,
Dreisow2011}. It was shown that in such systems the signal propagates along
a complex periodic trajectory. This phenomenon is known as a non-trivial, or
anharmonic Bloch oscillations.

The anharmonic Bloch oscillations are the subject of our work. In this paper
we generalize the results of \cite{Pertsch1998_Theor} to the case of the
zigzag arrays. We consider the system of equations of the CMM taking into
account the second order coupling. We find the analytical solution of this
system of equations and obtain the formula for the path of the optical beam.
The obtained formula allows to predict the geometric parameters of the
anharmonic Bloch oscillations, such as the period of the optical beam path
and the positions of the turning points. We demonstrate that the analytical
formula is consistent with the numerical calculations.


\section{Analytical derivation of the anharmonic Bloch oscillations by means
of the coupling modes model}

\label{Sec:AnalyticalDerivation}

Here we represent the analytical derivation of the optical beam path in a
zigzag array. Our derivation is based on the coupling mode equations with
second order coupling. Note that the calculation below is a generalization
of that represented in \cite{Pertsch1998_Theor}, where the ordinary Bloch
oscillations are considered.

We start from the equation of the coupling modes
\begin{equation}
\left(i\frac{d}{dz}+\beta_0^{(0)}+\alpha j\right)\,a_j(z)+ \gamma_1\Bigl(%
a_{j-1}(z)+a_{j+1}(z)\Bigr)+ \gamma_2\Bigl(a_{j-2}(z)+a_{j+2}(z)\Bigr)=0.
\label{AD005}
\end{equation}
Here
\begin{equation}
\beta_j^{(0)}=\beta_0^{(0)}+\alpha j,   \label{AD008}
\end{equation}
For the simplicity, we assume that $\beta_0^{(0)}=0$. Note that $%
\beta_0^{(0)}$ can be easily excluded from the equations by the replacement $%
a_j(z)=a^{\prime }_j(z)\,e^{i\beta^{(0)}_0 z}$.
\begin{equation}
\left(i\frac{d}{dz}+\alpha j\right)\,a_j(z)+ \gamma_1\Bigl(%
a_{j-1}(z)+a_{j+1}(z)\Bigr)+ \gamma_2\Bigl(a_{j-2}(z)+a_{j+2}(z)\Bigr)=0.
\label{AD010}
\end{equation}

We perform a Fourier transform of $a_j(z)$:
\begin{equation}
\tilde{a}(k,z)=\frac{1}{\sqrt{2\pi}} \sum\limits_j a_j(z)~e^{-ikj},
\label{AD020}
\end{equation}
\begin{equation}
a_j(z)=\frac{1}{\sqrt{2\pi}} \int\limits_{-\pi}^\pi dk~\tilde{a}%
(k,z)~e^{ikj}.   \label{AD030}
\end{equation}

From Eq. (\ref{AD010}) one can obtain an equation for $\tilde{a}(k,z)$:
\begin{equation}
\left(i\frac{d}{dz}+i\alpha\frac{d}{dk}\right)\,\tilde{a}(k,z)+ 2\Bigl(%
\gamma_1\,\cos\,k+\gamma_2\,\cos\,2k\Bigr)\,\tilde{a}(k,z)=0.   \label{AD040}
\end{equation}

Let us consider the eigenmode with a certain $\beta$:
\begin{equation}
\tilde{a}_\beta(k,z)=\tilde{a}_\beta(k)\,e^{i\beta z}.   \label{AD050}
\end{equation}
Substituting (\ref{AD050}) to (\ref{AD040}), one obtains:
\begin{equation}
\left(i\alpha\frac{d}{dk}-\beta+
2\gamma_1\,\cos\,k+2\gamma_2\,\cos\,2k\right)\,\tilde{a}_\beta(k)=0.
\label{AD060}
\end{equation}

Equation (\ref{AD060}) possesses the following solution:
\begin{equation}
\tilde{a}_\beta(k)=\exp\left\{\frac{-i}{\alpha} \Bigl(\beta
k-2\gamma_1\,\sin\,k-\gamma_2\,\sin\,2k\Bigr)\right\}.   \label{AD070}
\end{equation}

The function $\tilde{a}_\beta(k)$ has to be periodic in $k$. Thus, $%
\beta=\alpha n$, where $n$ is integer. The function $\tilde{a}_\beta(k)$ for
a certain $n$ is
\begin{equation}
\tilde{a}_n(k)=\exp\left\{ -ink+\frac{2i\gamma_1}{\alpha}\sin\,k+\frac{%
i\gamma_2}{\alpha}\sin\,2k \right\}.   \label{AD080}
\end{equation}

Now let us turn to the general solution of Eq. (\ref{AD040}). Any solution $%
\tilde{a}(k,z)$ can be represented as a superposition of eigenmodes (\ref%
{AD050}):
\begin{equation}
\tilde{a}(k,z)=\sum\limits_n C_n\,\tilde{a}_n(k)\,e^{i\alpha n z}.
\label{AD090}
\end{equation}

The coefficients $C_n$ can be obtained from the boundary conditions $\tilde{a%
}(k,z=0)=\tilde{a}^0(k)$. From (\ref{AD090}) it follows that
\begin{equation}
\begin{array}{c}
\displaystyle \tilde{a}^0(k)=\sum\limits_n C_n\,\tilde{a}_n(k)= \\
\displaystyle =\sum\limits_n C_n\,\exp\left\{ -ink+\frac{2i\gamma_1}{\alpha}%
\sin\,k+\frac{i\gamma_2}{\alpha}\sin\,2k \right\}.%
\end{array}
\label{AD100}
\end{equation}
Thus,
\begin{equation}
C_n=\int\limits_{-\pi}^\pi \frac{dk}{2\pi}~ \tilde{a}^0(k)~\exp\left\{ ink-%
\frac{2i\gamma_1}{\alpha}\sin\,k-\frac{i\gamma_2}{\alpha}\sin\,2k \right\}.
\label{AD110}
\end{equation}
Substituting (\ref{AD110}) to (\ref{AD100}), one obtains:
\begin{equation}
\begin{array}{c}
\displaystyle \tilde{a}(k,z)=\sum\limits_n \int\limits_{-\pi}^\pi \frac{%
dk^{\prime }}{2\pi}\, \tilde{a}^0(k^{\prime in(k^{\prime }-k+\alpha z)}\times
\\
\displaystyle \times\exp\left\{ \frac{2i\gamma_1}{\alpha}\Bigl(%
\sin\,k-\sin\,k^{\prime }\Bigr)+ \frac{i\gamma_2}{\alpha}\Bigl(%
\sin\,2k-\sin\,2k^{\prime }\Bigr)\right\}.%
\end{array}
\label{AD120}
\end{equation}
Since
\begin{equation}
\sum\limits_n e^{in(k^{\prime }-k+\alpha z)}=2\pi\,\delta(k^{\prime
}-k+\alpha z),  \label{AD130}
\end{equation}
Eq. (\ref{AD120}) takes the form:
\begin{equation}
\begin{array}{c}
\displaystyle \tilde{a}(k,z)=\tilde{a}^0(k-\alpha z)\times\, \\
\displaystyle \times\exp\left\{ \frac{2i\gamma_1}{\alpha}\Bigl(%
\sin\,k-\sin\,(k-\alpha z)\Bigr)+ \frac{i\gamma_2}{\alpha}\Bigl(%
\sin\,2k-\sin\,2(k-\alpha z)\Bigr)\right\}.%
\end{array}
\label{AD140}
\end{equation}

Let us assume that the function $\tilde{a}^0(k)$ possesses a sharp peak near
some $k_0$. Then, the function at the right-hand side of Eq. (\ref{AD140})
can be expanded into a Taylor series around $(k-\alpha z)-k_0$:
\begin{equation}
\tilde{a}(k,z)=\tilde{a}^0(k-\alpha z)~ e^{i\phi(z)+i(k-\alpha
z-k_0)\,\psi(z)}   \label{AD150}
\end{equation}
where
\begin{equation}
\phi(z)= \frac{2\gamma_1}{\alpha}\Bigl(\sin\,(k_0+\alpha z)-\sin\,k_0\Bigr)+
\frac{\gamma_2}{\alpha}\Bigl(\sin\,2(k_0+\alpha z)-\sin\,2k_0\Bigr),
\label{AD160}
\end{equation}
\begin{equation}
\psi(z)= \frac{2\gamma_1}{\alpha}\Bigl(\cos\,(k_0+\alpha z)-\cos\,k_0\Bigr)+
\frac{2\gamma_2}{\alpha}\Bigl(\cos\,2(k_0+\alpha z)-\cos\,2k_0\Bigr).
\label{AD170}
\end{equation}

Substitute (\ref{AD150}) to (\ref{AD030}):
\begin{equation}
a_j(z)=e^{i\phi(z)+i(k_0+\alpha z)j} \int\limits_{-\pi}^\pi \frac{dk}{\sqrt{%
2\pi}}\, \tilde{a}^0(k-\alpha z)\,e^{i(k-\alpha z-k_0)(j+\psi(z))}.
\label{AD180}
\end{equation}

The intensity of the optical beam at the $j$-th waveguide is defined as $%
I_j(z)=|a_j(z)|^2$. Thus,
\begin{equation}
I_j(z)= \left|\int\limits_{-\pi-k_0-\alpha z}^{\pi-k_0-\alpha z} \frac{dk_1}{%
\sqrt{2\pi}}\,\tilde{a}^0(k_0+k_1)\, e^{ik_1(j+\psi(z))}\right|^2,
\label{AD190}
\end{equation}
where $k_1=k-\alpha z-k_0$.

Since $\tilde{a}^0(k)$ has a sharp peak at $k=k_0$, the dependence of the
integral limits on $z$ does not affect the integral value. So, we can write
\begin{equation}
I_j(z)=I(j+\psi(z))   \label{AD200}
\end{equation}
Thus, the trajectory of the optical beam is defined with the equation $%
j+\psi(z)=\mathrm{const}$. This equation results in
\begin{equation}
j(z)=j_0- \frac{2\gamma_1}{\alpha}\Bigl(\cos\,(k_0+\alpha z)-\cos\,k_0\Bigr)%
- \frac{2\gamma_2}{\alpha}\Bigl(\cos\,2(k_0+\alpha z)-\cos\,2k_0\Bigr).
\label{AD210}
\end{equation}

Eq. (\ref{AD210}) represents the trajectory of the optical beam in the
zigzag array of waveguides.


\section{Qualitative analysis of the analytical formula}

\label{Sec:QualitativeAnalysis}

Let us analyze the properties of the trajectory described with Eq. (\ref%
{AD210}). For the simplicity, we consider the special case $j_0=0$ and $k_0=0
$. For this case Eq. (\ref{AD210}) takes the form
\begin{equation}
j(z)=\frac{2\gamma_1}{\alpha}\,\Bigl(1-\cos\,\alpha z\Bigr) +\frac{2\gamma_2%
}{\alpha}\,\Bigl(1-\cos\,2\alpha z\Bigr).  \label{QA010}
\end{equation}
It is obvious that $j(z)$ is the periodical function with the period $%
2\pi/\alpha$. Below we consider the form of the trajectory in a single
period, $0\leq z<2\pi/\alpha$. We assume that the coupling constants $%
\gamma_1$ and $\gamma_2$ are of the same sign (both positive or both
negative). This assumption is consistent with experiments.

Let us find the turning points of the trajectory. For this purpose we have
to solve the equation
\begin{equation}
\frac{dj}{dz}(z)= 2\gamma_1\,\sin\,\alpha z+4\gamma_2\,\sin\,2\alpha z=
2\gamma_1\,\sin\,\alpha z\, \left(1+\frac{4\gamma_2}{\gamma_1}\,\cos\,\alpha
z\right)=0.   \label{QA020}
\end{equation}
It follows from (\ref{QA020}), that the turning points are defined with two
equations:
\begin{equation}
\sin\,\alpha z=0,   \label{QA030a}
\end{equation}
\begin{equation}
\cos\,\alpha z=-\frac{\gamma_1}{4\gamma_2}.   \label{QA030b}
\end{equation}

We should consider two different cases, namely $\gamma_1/4\gamma_2>1$ and $%
\gamma_1/4\gamma_2<1$.

\begin{figure}[tbp]
\includegraphics[width=0.4\textwidth]{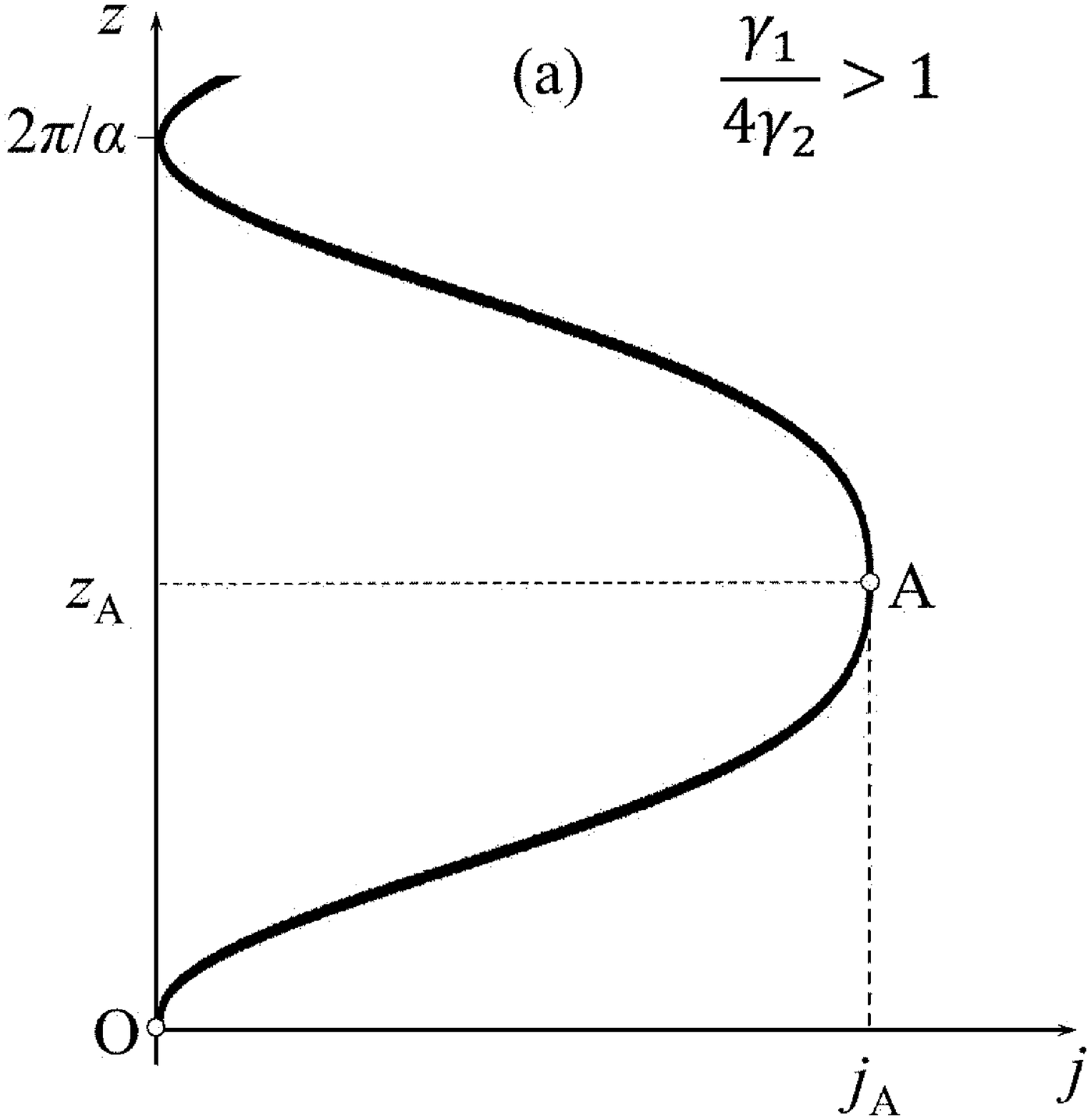} \hfill %
\includegraphics[width=0.4\textwidth]{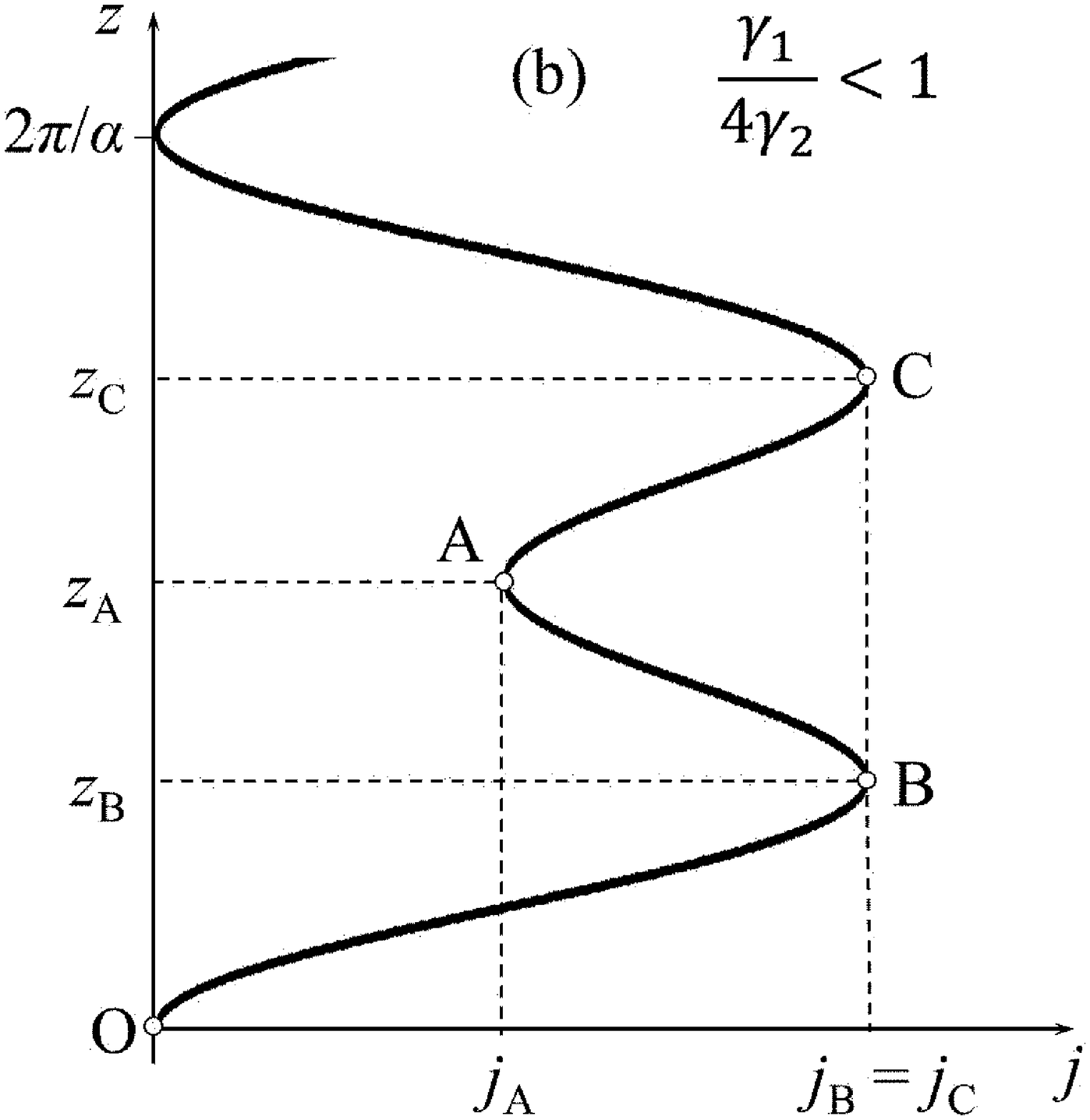} \newline
\caption{ Schematic image of the trajectory described by Eq. ({\protect\ref%
{QA010}}) for two different cases: \newline
(a) $\protect\gamma_1/4\protect\gamma_2>1$; $\qquad$ (b) $\protect\gamma_1/4%
\protect\gamma_2<1$. }
\label{Fig_SchemeTraj}
\end{figure}

Let us begin with the case $\gamma_1/4\gamma_2>1$. For this case Eq. (\ref%
{QA030b}) has no solutions, and all the turning points are defined with Eq. (%
\ref{QA030a}) only. In a single period $0\leq z<2\pi/\alpha$ Eq. (\ref%
{QA030a}) possesses two solutions, $z_O=0$ and $z_A=\pi/\alpha$. Thus, in
one period of the trajectory there are two turning points. The schematic
image of the trajectory is represented in Fig. \ref{Fig_SchemeTraj}(a). In
this figure, the turning points are designated with the letters O and A.
Substituting the values $z_O$ and $z_A$ to Eq. (\ref{QA010}), one can find
the number $j$ of a waveguide where the trajectory changes the direction: $%
j_O=0$ and $j_A=4\gamma_1/\alpha$.

Now let us turn to the other case, $\gamma_1/4\gamma_2<1$. The schematic
image of the trajectory for this case is represented in Fig. \ref%
{Fig_SchemeTraj}(b). For this case, the both equations (\ref{QA030a}) and (%
\ref{QA030b}) possess the solutions. Thus, in a period $0\leq z<2\pi/\alpha$
the trajectory possesses four turning points. Two of them are the solutions
of (\ref{QA030a}), they are designated as O and A, and their coordinates $%
\{z_O,\,j_O\}$ and $\{z_A,\,j_A\}$ are obtained above. Two other turning
points are denoted with B and C. Their coordinates $z$ are the solutions of
Eq. (\ref{QA030b}): $\displaystyle z_B=\frac{1}{\alpha}\left(\pi-\arccos\,%
\frac{\gamma_1}{4\gamma_2}\right)$ and $\displaystyle z_C=\frac{1}{\alpha}%
\left(\pi+\arccos\,\frac{\gamma_1}{4\gamma_2}\right)$. Substituting $z_B$
and $z_C$ to (\ref{QA010}), one obtains $\displaystyle j_B=j_C=\frac{%
4\gamma_2}{\alpha}\,\left(1+\frac{\gamma_1}{4\gamma_2}\right)^2$.


\section{Comparison of the analytical formula with the numerical calculation}

\label{Sec:Comparison}

Below we compare the trajectory described with the analytical formula (\ref%
{AD210}) with the numerical calculation based on Eq. (\ref{AD005}). For this
purpose we take the same parameters as in the paper \cite{Wang2010}. The
calculations are produced for $\alpha=0.2$, $\gamma_1=1$, and for two values
of the parameter $\gamma_2=0$ and $0.7$.

For the numerical solution of Eq. (\ref{AD005}) we have to specify some
boundary conditions $a_j(z)$ for $z=0$. We take the boundary conditions in
the form of the Gaussian beam
\begin{equation}
a_j(0)=e^{-\frac{(j-j_0)^2}{\sigma^2}},   \label{Co010}
\end{equation}
where $j_0=10$ and $\sigma=4$. The Fourier transform $\tilde{a}^0(k)=\tilde{a%
}(k,z=0)$ defined by Eq. (\ref{AD020}) takes the form
\begin{equation}
\tilde{a}^0(k)=e^{-\frac{\sigma^2 k^2}{4}}.   \label{Co020}
\end{equation}
This function possesses a sharp peak near $k=0$. Thus, in the analytical
formula (\ref{AD210}) we take $k_0=0$.

\begin{figure}[tbp]
\includegraphics[width=0.45\textwidth]{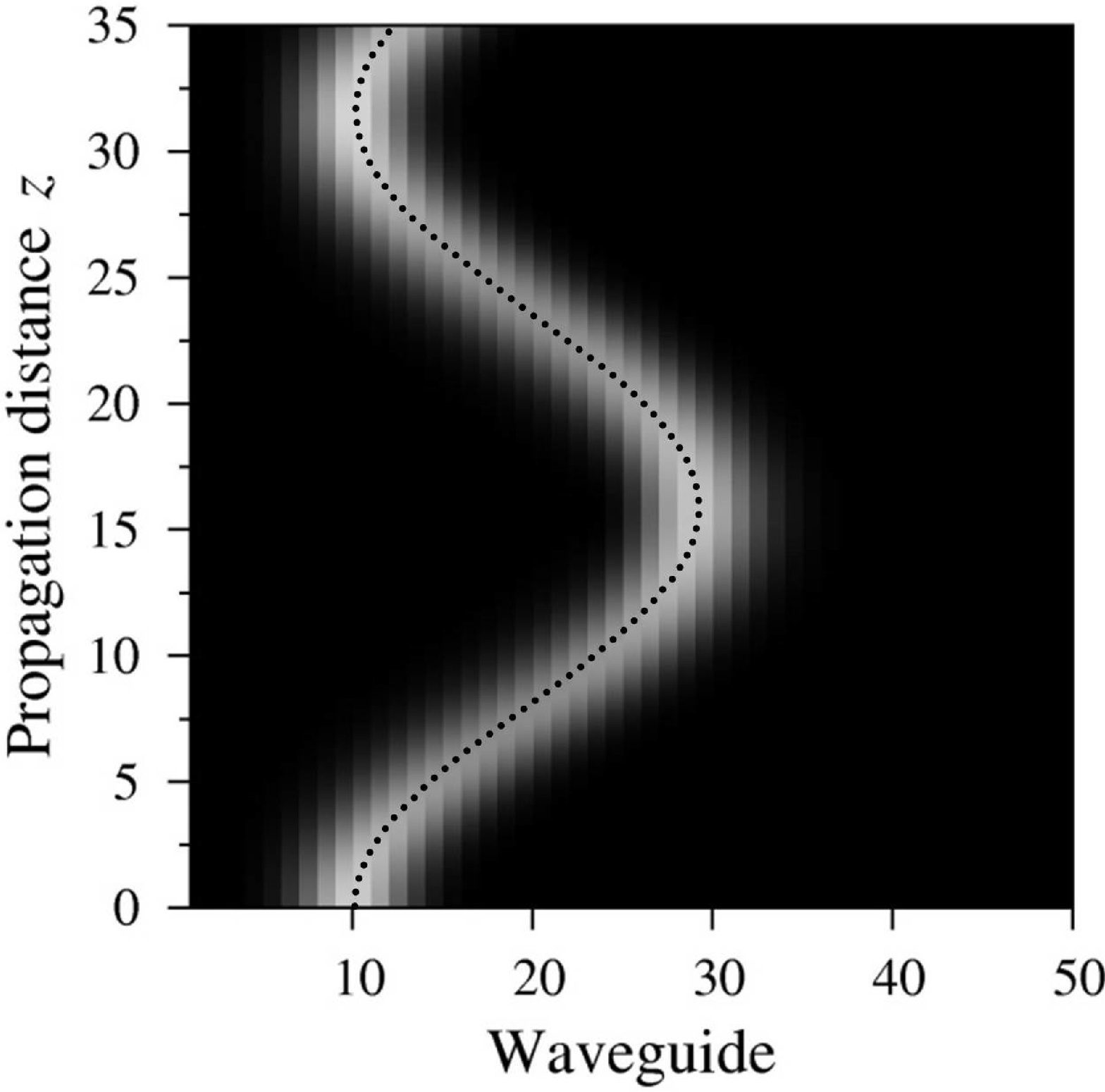} \hfill %
\includegraphics[width=0.45\textwidth]{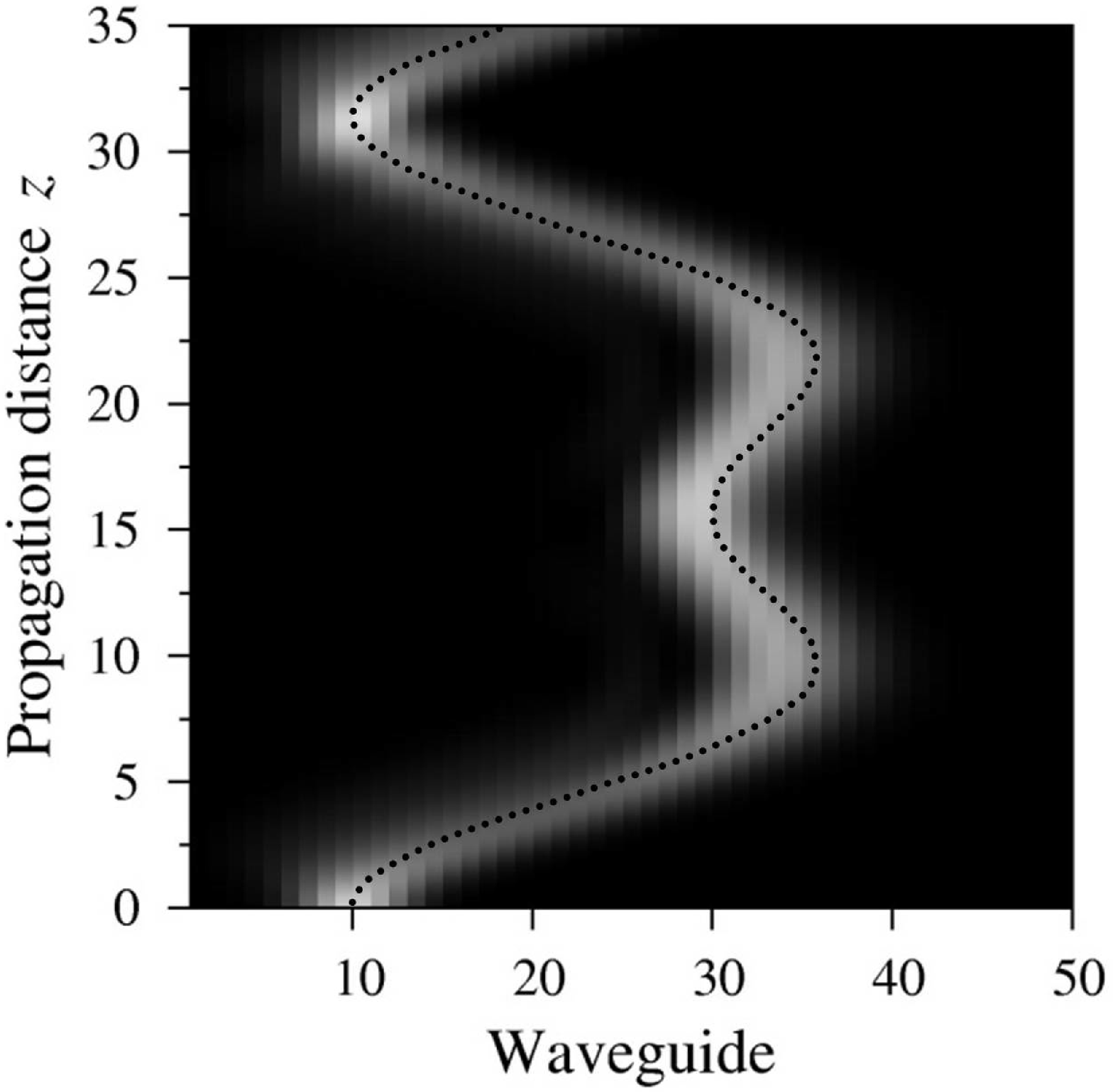} \newline
\caption{ The trajectory of the optical beam. Left: $\protect\gamma_2=0$;
right: $\protect\gamma_2=0.7$. }
\label{Fig_NumSim}
\end{figure}

The results of our calculations are represented in Fig. \ref{Fig_NumSim}.
The white trail illustrates the Gaussian beam path calculated numerically by
means of Eq. (\ref{AD005}). The dotted line is the trajectory calculated
with the analytical formula (\ref{AD210}). The comparison of the numerical
and analytical calculations confirms that the analytical formula describes
the trajectory of the Gaussian beam exactly.

\section{Conclusion}

\label{Sec:Conclusion}

In this paper we investigated the anharmonic Bloch oscillations by means of
the coupling modes model. For this problem, the equations of the CMM can be
solved analytically. This circumstance allowed us to obtain the analytical
expression for the path of the optical beam.

However, to use the obtained formula, one has to define the parameters of
the CMM, namely, propagation constants and coupling constants. Typically,
the values of these parameters are obtained experimentally.

Recently we proposed a method to calculate the parameters of the
CMM by means of the multiple scattering formalism \cite{We_OptEng2014}. This
allows to simplify significantly the investigation of the optical Bloch
oscillations.

Besides, let us notice that the CMM is approximate model, and can be used
only for the weak interaction between the waveguides. Otherwise, one should
use the exact methods of numerical simulation. In this case, the analytical
formula obtained in this work may be inapplicable.

\bigskip

\subsection*{Acknowledgments.}

The study is supported by the Russian Fund for Basic Research (Grants
13-02-00472a and 14-29-08165) and by the Ministry of Education and Science
of Russian Federation, project 8364.


\section{Conclusion}

\end{document}